\documentclass{article}

\usepackage[utf8]{inputenc}
\usepackage[left=2cm,right=2cm,top=2cm,bottom=2cm]{geometry}
\usepackage[utf8]{inputenc}
\usepackage[T1]{fontenc}
\usepackage{bm}
\usepackage{xcolor}
\usepackage{graphicx}
\usepackage{url}
\usepackage{authblk}

\newcommand{\exciting}{{\usefont{T1}{lmtt}{b}{n}exciting}}
\newcommand{\eg}{{\it e.g.}, }
\newcommand{\ie}{{\it i.e.}, }
\newcommand{\fsub}[1]{$_\mathrm{#1}$}

\title{Similarity of materials and data-quality assessment by fingerprinting}

\author[1]{Martin Kuban}
\author[1]{\v{S}imon Gabaj}
\author[1]{Wahib Aggoune}
\author[1]{Cecilia Vona}
\author[1]{Santiago Rigamonti} 
\author[1, *]{Claudia Draxl}
\affil[1]{Institut f\"ur Physik and IRIS Adlershof, Humboldt-Universit{\"a}t zu Berlin, 12489 Berlin, Germany}
\affil[*]{Corresponding author: draxl@physik.hu-berlin.de}
\date{January 2022}

\begin{document}

\maketitle

\begin{abstract}
Identifying {\it similar} materials, \ie those sharing a certain property or feature, requires {\it interoperable} data of high quality. It also requires means to measure similarity. We demonstrate how a \textit{spectral fingerprint} as a descriptor, combined with a similarity metric, can be used for establishing quantitative relationships between materials data, thereby serving multiple purposes. This concerns, for instance, the identification of materials exhibiting electronic properties similar to a chosen one. The same approach can be used for assessing uncertainty in data that potentially come from different sources. Selected examples show how to quantify differences between measured optical spectra or the impact of methodology and computational parameters on calculated properties, like the the density of states or excitonic spectra. Moreover, combining the same fingerprint with a clustering approach allows us to explore materials spaces in view of finding (un)expected trends or patterns. In all cases, we provide physical reasoning behind the findings of the automatized assessment of data. 
\end{abstract}

\section{Impact statement}
To predict novel materials with desired properties, data-centric approaches are in the process of becoming an additional fundament of materials research. Prerequisite for their success are well-curated data. Ideally, one can make use of multiple data collections. Bringing data from different sources together, poses challenges on their {\it interoperability}, which are routed in two out of the {\it 4V} of Big Data. These are the uncertainty of data quality (\textit{veracity}) and the heterogeneity in form and meaning of the data (\textit{variety}). To overcome this barrier, universal and interpretable measures must be established which quantify differences between data that are supposed to have the same meaning. Here, we show how a spectral fingerprint in combination with a similarity metric can be used for assessing spectral properties of materials. Our approach allows for tracing back in computed as well as measured data, differences stemming from various aspects. It thus paves the way for automatized data-quality assessment towards interoperability. Based on this, in turn, materials exhibiting similar features can be identified.

\section{Introduction}
\label{sec:introduction}
Finding materials for a specific application, may be a long and tedious process. Typical time spans from the scientific invention to the market are twenty years or even longer \cite{MGI}. Thus, there is urgent demand for speeding up the process to identify candidate materials that exhibit desired properties. This is particularly so in view of the enormous challenges arising from the world's tremendously increasing energy consumption and environmental problems. Overall, there is hardly any part of our society that is not concerned with materials, and the materials-science community is dealing with a wide variety of materials classes, their properties and functions. 

For identifying materials with desired properties, data-centric approaches \cite{MRS} are attracting a lot of attention. For being successful on a large scale, a fundamental requirement is to make use of the data collections from the entire community. Indeed, more and more publicly available data collections are being established, and efforts towards "FAIRification" \cite{Wilkinson2016} are growing world-wide. This raises in particular the question of interoperability, the $I$ in FAIR. Bringing together data from various sources implies heterogeneity. In other words, {\it variety} and {\it veracity}, two of the {\it 4V} of Big Data are becoming an issue. Therefore, it is crucial to assess and control the uncertainty in data \cite{Scheffler2022}. On the experimental side, {\it variety} concerns different methods to measure a physical property, and -- within any such method -- diverse instruments together with a possible selection of measurement modes. Likewise, computing a specific property, can be done by different methodologies and possible approximations, utilizing different software packages. Uncertainty in computed data, in turn, is related to algorithms, implementation, and computational parameters. These correspond to different resolution and measurement conditions (\eg pressure, temperature, environment) on the experimental side. The latter is also concerned with the quality and growth condition or treatment of the materials sample. All these uncertainties need to be understood, and ideally be quantified, to allow for unrestricted interoperability. Obviously, this is an overwhelming task. It also illustrates the urgent need for benchmark data to quantify deviations from the ideal results. For the computational materials side, the topics of reproducibility \cite{Lejaeghere2016} and benchmarks for solids \cite{Gulans2018,Jensen2017,Nabok2016,Rangel2020} are being pursued since a few years only.

For the identification of materials with specific features, similarity is an important concept. Materials of interest for a given application should share specific properties, \ie being similar in some aspects while they may be very different in others. To assess similarity in general and, in view of data-centric approaches in particular, one needs to introduce adequate descriptors and similarity measures \cite{Ramprasad2017,Isayev2015,Mahmoud2020,Gjerding_2021,Knosgaard2022,kuban2022} that go beyond that of the atomic structure \cite{De2016}. 

In this work, we show how both aspects -- identifying similar materials and quantifying uncertainties -- can be addressed with methods and tools measuring similarity. More specific, we assess similarity in electronic properties in terms of a spectral fingerprint. We demonstrate our approach by various scenarios like identifying materials that are similar to a chosen one, measuring the impact of structural features as well as methodology or computational parameters on the accuracy and precision of computed properties, or highlighting differences in sample quality or measurement details on experimental results. The same descriptor can also be combined with unsupervised learning to identify and analyze trends in large data sets as demonstrated recently \cite{kuban2022}. In all our examples, we provide physical reasoning behind the data-based observations. Finally, we discuss how this approach can be used to enhance large-scale data collections.

\section{Results}
The following examples focus on exploring and understanding data spaces on the one hand, and on highlighting effects that could potentially lead to {\it veracity} on the other hand. We emphasize that all calculations shown here are perfectly valid but may differ in certain numerical aspects as each of them has been created for a specific purpose. The examples are neither chosen such to represent the best possible calculations nor to showcase inconsistencies, but only to highlight the respective scenario that we like to demonstrate. They help to illustrate where differences between data that should mean or describe the same, may come from and how these differences can be quantified. This quantification is carried out by combining the spectral fingerprint with the Tanimoto coefficient (Tc) as a similarity measure \cite{Willett1998}, as described in Section \ref{sec:methods}. In short, Tc varies between 0 (completely different) to 1 (identical). 

\begin{figure}[hbt]
    \centering
    \includegraphics[width = 0.7\textwidth]{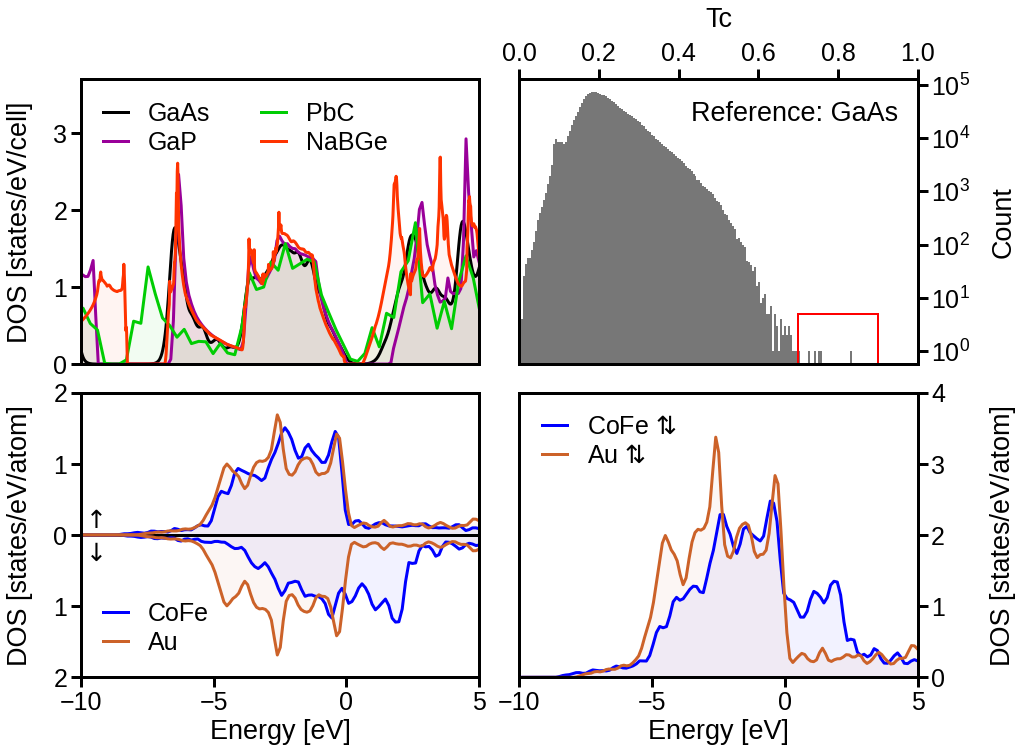}
    \caption{DOS of similar materials from a search within 1.8 million materials in the NOMAD Encyclopedia \cite{NOMAD2018,NOMAD2019}. The top left panel shows the three materials most similar to the semiconductor GaAs, \ie GaP ($\mathrm{Tc} = 0.83$), PbC ($\mathrm{Tc} = 0.75$), and NaBGe ($\mathrm{Tc} = 0.74$). The top right panel represents the distribution of Tc values of the reference with all considered materials (note the logarithmic scale). The red box indicates the most similar materials. The lower panels depict the metals Au and CoFe which share high similarity in the majority spin ($\mathrm{Tc} = 0.80$), but owing to the very different minority spin channels ($\mathrm{Tc} = 0.47$), the overall $\mathrm{Tc}$ is 0.67 only.}
    \label{fig:similar_materials}
\end{figure}

\subsection{Identifying similar materials}
\label{sec:similar}

Our first case is about finding materials that share a certain characteristics with a selected reference. Replacing, for instance, toxic or rare elements or components with less harmful or abundant ones, is one important aspect in the search of new materials. Here, we demonstrate how to find materials that share features of their electronic structure. Using the spectral fingerprint (see Methods below), we  have searched about 1.8 million materials in the NOMAD Encyclopedia (corresponding to about $93\, \%$ of the materials available today) in order to find materials that have a similar electronic density of states (DOS). Figure \ref{fig:similar_materials} presents two selected results of this investigation. The top left panel shows the most similar materials to the semiconductor GaAs. We observe that all four compounds share the higher-lying valence band structure and differ either in the lower valence region or the conduction bands. As could be expected from the valence configurations of atoms from the same column of the periodic table of elements (PTE), GaP -- also sharing the same crystal structure -- is a good candidate that indeed, turns out to be the material most similar to GaAs ($\mathrm{Tc} = 0.83$). Maybe less expected are PbC ($\mathrm{Tc} = 0.75$) and NaBGe ($\mathrm{Tc} = 0.74$). To obtain a better understanding about what the observed Tc values mean, we provide a histogram (top right) that shows the distribution of Tanimoto coefficients of the reference DOS with that of all 1.8 million materials. Strikingly, materials with high similarity coefficients are exceptional. The red box indicates those with $\mathrm{Tc} > 0.7$. We note, however, that the distribution depends on the considered energy range (here from -10 eV to 5 eV). Choosing a narrower energy window to focus on specific features of the spectra, may lead to a larger number of materials  that are similar to the reference.

The lower panels of Fig. \ref{fig:similar_materials} demonstrate how (dis)similar the two metals Au and CoFe are. Here, the spin-properties govern the behavior. The majority spin exhibits a high similarity of $\mathrm{Tc} = 0.80$, where the DOS reflects the akin character of the occupied 5$d$ Au and 3$d$ bands of Cu, respectively. In contrast, the minority spins ($\mathrm{Tc} = 0.47$) differ mainly due to a rigid shift of the partially filled  Cu 3$d$ minority band by about $2.5\, \mathrm{eV}$ such that the overall similarity is only moderate ($\mathrm{Tc} = 0.67$).

\subsection{Impact of methodology}
\label{sec:methodology}

\begin{figure}[hbt]
    \centering
    \includegraphics[width = 0.7\textwidth]{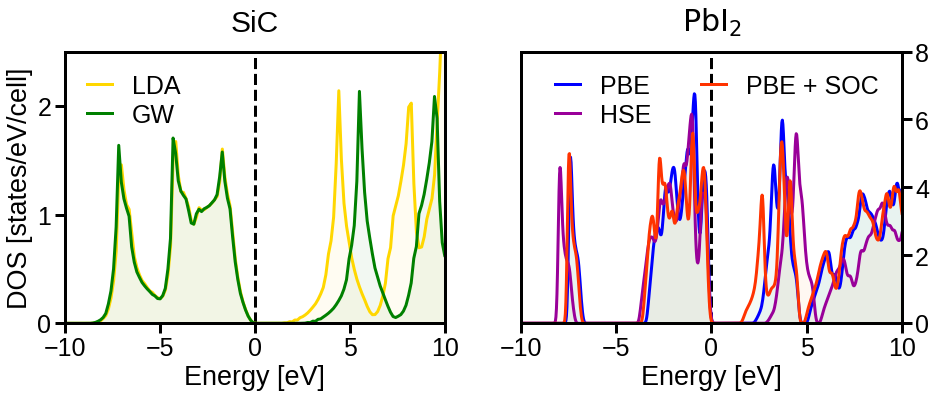}
    \caption{DOS of SiC \cite{NOMADSiCGW} (left panel), calculated with the all-electron code \exciting\ by employing the LDA (yellow) and the $G_0W_0$ approach (green). The overall similarity coefficient is $\mathrm{Tc} = 0.66$. While the valence regions are nearly indistinguishable ($\mathrm{Tc} = 0.96$), the conduction bands, with a small $\mathrm{Tc}$ of $0.27$, show the well-known upward shift by $G_0W_0$. The right panel demonstrates the strong impact of exchange-correlation (xc) functional (PBE vs HSE06, $\mathrm{Tc} = 0.60$) and spin-orbit coupling (SOC) (PBE vs PBE+SOC, $\mathrm{Tc} = 0.71$) in the case of PbI$_2$ \cite{Vona2021}. The dashed vertical lines indicate the Fermi energy.}
    \label{fig:LDA_GW}
\end{figure}

Figure \ref{fig:LDA_GW} compares the density of states of SiC \cite{NOMADSiCGW}, calculated by the local-density approximation (LDA) and the $G_0W_0$ approximation of many-body perturbation theory. Both results are obtained with the \exciting\ code \cite{Gulans2014}, using the same set of computational parameters. Hence, the difference in the two results can be solely assigned to the method. Only assessing the valence-band region, gives a similarity coefficient of $\mathrm{Tc} = 0.96$. In contrast, the conduction region reveals the well-known opening of the electronic band gap by $G_0W_0$, leading to a small similarity coefficient of $0.27$ only. The overall similarity covering the entire energy range is moderate, \ie $\mathrm{Tc} = 0.66$.

The panel on the right draws a similar picture for PbI$_2$. Here, we explore on the one hand how accounting for spin-orbit coupling (SOC) changes the results. In essence, by decreasing the band gap by as much as 0.68 eV, SOC showing up in a modest Tanimoto coefficient of $\mathrm{Tc} = 0.71$ when considering the entire energy range of $-10$ to $10\, \mathrm{eV}$. On the other hand, going from PBE to HSE06, the gap opens by 0.69 eV, blue-shifting the DOS by this amount and leading, with PBE as the counterpart, to $\mathrm{Tc} = 0.60$. Overall, the competition of the two effects, giving rise to cancelation of errors, explains why the Kohn-Sham gap obtained by PBE without SOC is rather close to experiment \cite{Gahailler1969,Ahuja2002,Shen2018}.
Considering the valence region only, both SOC and HSE have a significant impact, giving rise to Tc values of 0.75 (PBE vs PBE+ SOC) and 0.73 (PBE vs HSE). The conduction region is characterized by an upward shift of all bands by HSE ($\mathrm{Tc} = 0.45$ for PBE vs. HSE), while SOC only affects the bands up to 5 eV ($\mathrm{Tc} = 0.67$ for PBE vs PBE+SOC), thus changing the shape of the DOS.

Obviously, these examples confirm in an automated way what we usually observe by visual inspection. At the same time, it illustrates a workflow that can be applied to large datasets, when visual inspection becomes unfeasible. Given very many calculations for many materials, our fingerprint will eventually allow us to extract knowledge that we cannot obtain from individual calculations or publications. The longer-term goal here is to verify such observations on large datasets, and provide and implement automated tools to simplify the here presented analysis. This will enable us to detect, on a large scale, for which material class what level of methodology is needed to provide a reliable result. In other words, we will learn what one can expect from the performance of a method for a given material or property, and what we can recommend to a novice user. In the examples below, we will take the idea further to include also computational parameters in these considerations.

\subsection{Excitonic spectra}
\label{sec:excitons}

\begin{figure}[h]
    \centering
       \includegraphics[width = 0.95\textwidth]{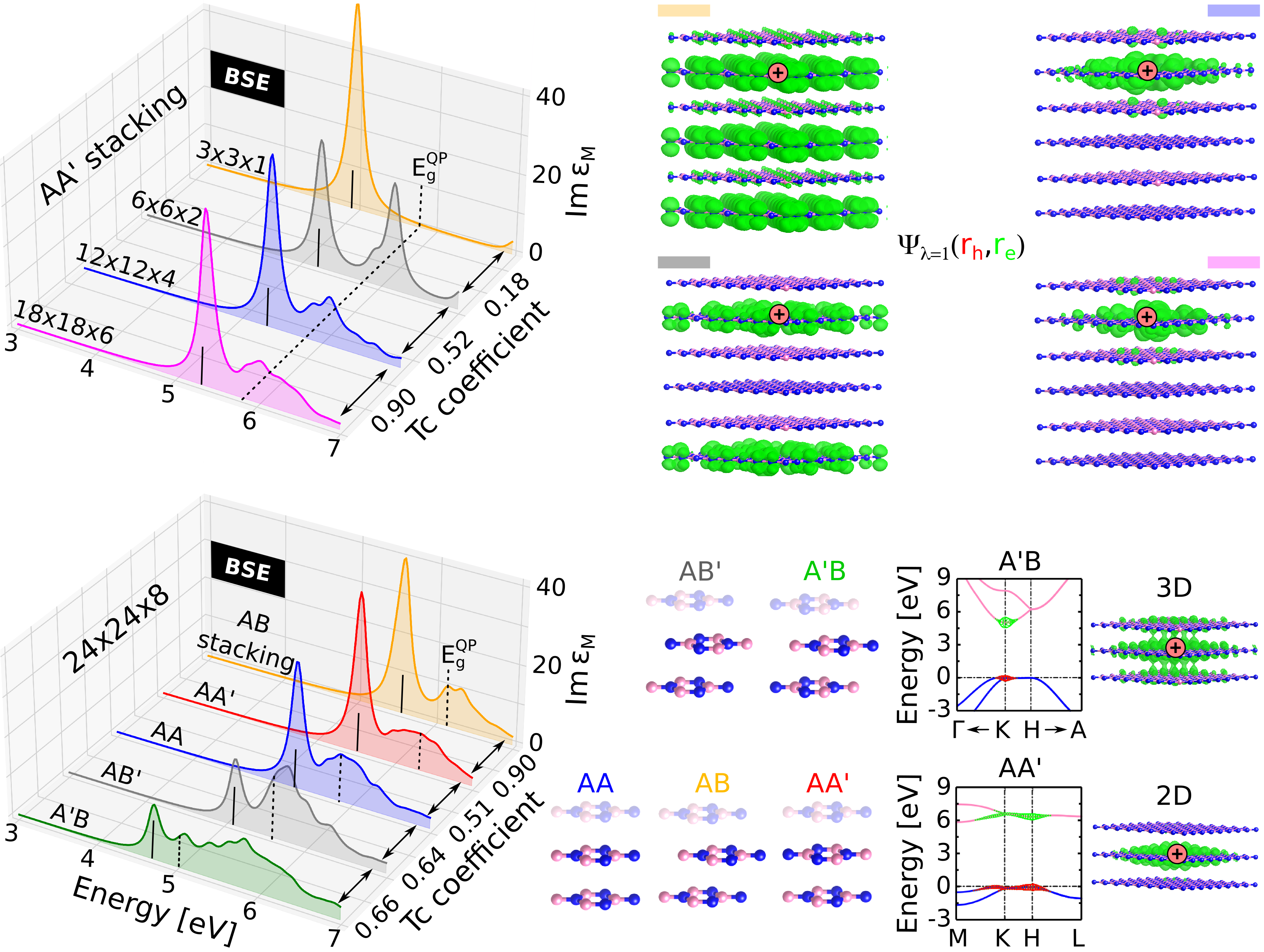}
    \caption{Optical spectra of h-BN. Top left: In-plane component of the dielectric tensor computed with different {\bf k}-point grids where the similarity coefficient between two adjacent curves is indicated on the right. The vertical dashed and solid lines represent the direct quasi-particle gaps and absorption onsets, respectively. The electron-hole wavefunctions in the top right show a changing degree of localization (note the main text). The electron density distribution is shown for the hole position fixed at the red dot. Bottom left: Converged spectra for different stacking motifs shown next to the graph. On the bottom right, the contributions from valence and conduction bands involved in the lowest-energy exciton for two representative stackings are displayed together with the corresponding exciton wavefunctions.}
    \label{fig:h-BN_spectra}
\end{figure}
In Fig. \ref{fig:h-BN_spectra}, we display how the optical spectra of h-BN in A'A stacking evolve as a function of the {\bf k}-grid used in the calculations. The spectra are characterized by an absorption onset (solid vertical lines) far below the band gap (dashed line), indicative of strong excitonic effects \cite{Aggoune2018}. The excitonic binding energy decreases with improved sampling, \ie the onset experiencing a blue-shift. This seems in contradiction to the corresponding exciton wavefunctions displayed on the right that one would expect to become more delocalized. The latter is indeed the case. Looking, for instance, at the behavior in {\bf c} direction, the reason for having the same electron distribution in every second plane is simply explained by the fact that the unit cell contains two such planes and, restricting ourselves to 1 {\bf k}-point in this direction, we obtain replica of the same wavefunction in every other plane. In other words, the exciton is spatially {\it not} much delocalized. The same applies to the in-plane directions where we also observe a seeming delocalization. Only by considering a denser {\bf k}-grid, the localized character becomes apparent. 

Making use of the fingerprinting in the exact same manner as for the DOS, one can {\it measure} the convergence behavior. The similarity coefficient between the $3\times3\times1$ and  $6\times6\times2$ results is 0.18, increasing to 0.52 between  $6\times6\times2$, and $12\times12\times4$ and to 0.90 between $12\times12\times4$ and  $18\times18\times6$. The latter two spectra are spot on in the onset region and only slightly differ at higher energies, causing the deviations from the optimal Tc value (ideally being 1) as it is measured over the entire energy range. The gradual increase in similarity indeed reflects the behavior of the exciton wavefunction. On the longer run, based on a large-enough data pool, we expect to learn optimal parameter settings for a given material class such to provide recommendations to users who want to carry out such computationally expensive calculations with high numerical precision at least possible effort. 

The lower half of Fig. \ref{fig:h-BN_spectra} shows the impact of structural features on the optical spectra of h-BN. On the left, the BSE spectra of five different stackings are shown that are displayed next to the figure. From a first inspection, we can discriminate between two groups: (i) The spectra of the AB, AA' and AA stackings have similar shape and intensity while (ii) those of AB' and A'B exhibit lower-energy excitations with smaller oscillator strength. As evident from the exemplary exciton wavefunction on the right, the intense peak at the absorption onset of the first group is characterized by a localized exciton with 2D distribution. In these cases, the degeneracy of the valence- and conduction-band edges is governed by how the B and N atoms are aligned in the out-of-plane direction~\cite{Aggoune2018}. The resulting character of the initial and final states of the excitation leads to its 2D distribution~\cite{Aggoune2018}. The similarity coefficient between AB and AA' results is 0.9 and only 0.51 between AA' and AA. Despite the similar shape of the spectra, the absorption onset of the latter is lower in energy due to its smaller quasi-particle gap (dashed lines). The moderate similarity coefficients between the spectra of the first and the second group reflects the differences in their band structure. The valence bands (conduction bands) in the AB' (A'B) stackings are split, which is a consequence of the vertical alignment of atoms of the same species. This impacts the exciton distribution which is of 3D character in these cases and exhibit lower binding energy compared to the first group. Overall, this example shows the direct relationship between structural arrangement, band structure, absorption spectrum, and exciton character. The similarity coefficients are able to capture (or measure) the main differences between various results. On the longer term, based on the analysis of many spectra, one may be able to provide recommendations which configurations may generate bound excitons with localized/delocalized character.

\subsection{Interoperability of experimental spectra}
\label{sec:experiment}

\begin{figure}[hbt]
    \centering
    \includegraphics[width = 0.7\textwidth]{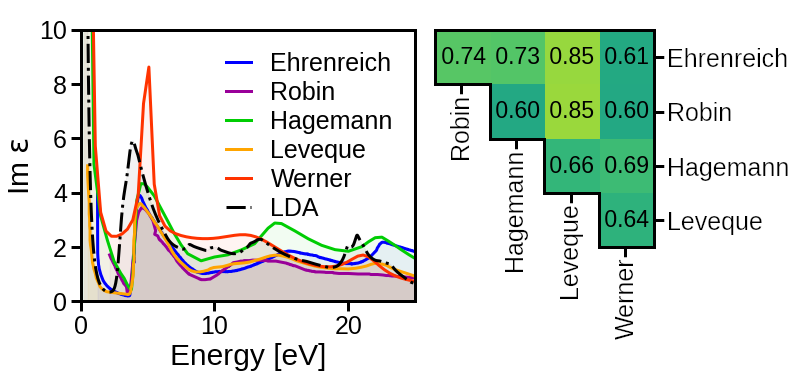}
        \caption{Optical spectra of the elemental solid silver from different sources and measurements \cite{Ehrenreich1962,Robin1966,Hagemann1975,Leveque1983, Werner2009}. The background of the similarity matrix on the right shows the same color code as Fig. \ref{fig:codes}. Due to symmetry, only the upper off-diagonal elements of the matrix are shown. For comparison, a calculation within the independent-particle approximation based on LDA is shown. Its Tc value with the spectra from Werner {\it et al.} is 0.78, with that of Robin, it is 0.68 only. }
    \label{fig:Al_experiment}
\end{figure}

Interoperability of experimental data is even a bigger issue compared to computational results. For instance, one and the same physical property can be measured by different methods, giving rise to {\it variety} in the data. The dielectric function is a good example for this, as it can be obtained by several experimental probes, being optical absorption or reflection spectroscopy, ellipsometry, as well as electron-loss spectroscopy. None of the measurements yields this property directly but only after some transformation steps or modeling behind, which adds a {\it veracity} problem. Overall, as pointed out in the Introduction, a measurement may depend on a huge number of parameters that need to be captured by metadata.

So, we may ask whether it is surprising or not that the optical spectra of silver displayed in Fig. \ref{fig:Al_experiment} exhibit so low similarity among each other. In the ideal world, one would expect spectra of an elemental, noble metal to be spot on. Reality tells us that this is not the case. While the measurements displayed here, all agree on the absorption onset of interband transitions, the peak positions at higher energies differ up to several eV, and the oscillator strengths are within roughly a factor of two. Without further information, it is impossible to tell which of the measurements may be superior to others and why. Obviously, both sample quality as well as measurement conditions have a strong impact, and a full annotation of the measured data with metadata are key for a fair comparison of experimental data. 

{\it Veracity} is also an issue when comparing measured and calculated data. It is needless to say that the dielectric function can be computed with various theoretical methods of different levels of sophistication. Here, we show an LDA result within the independent-particle approximation \cite{Werner2009}. Measuring its similarity to the experimental data of Robin  \cite{Robin1966} and Werner {\it et al.} \cite{Werner2009}, delivers Tc values of {$0.63$ and $0.78$}, respectively. The better agreement with the latter -- the most recent of the experimental datasets -- has been discussed elsewhere \cite{Werner2009}. We note that, to a large extent, the disagreement comes from the underestimated absorption onset by LDA, owing to the too high position of the Ag-$d$ bands.

This example shows how heterogeneous experimental data of even very simple and well understood materials can be. This situation imposes the need for detailed annotation of measured data, and it also highlights the urgent need for a large variety of benchmark results that serve as reference data. Otherwise, an in-depth assessment of data quality on a large scale will remain difficult.

\subsection{Influence of code and computational parameters}
\label{sec:parameters}

\begin{figure}[hbt]
    \centering
    \includegraphics[width = 0.85\textwidth]{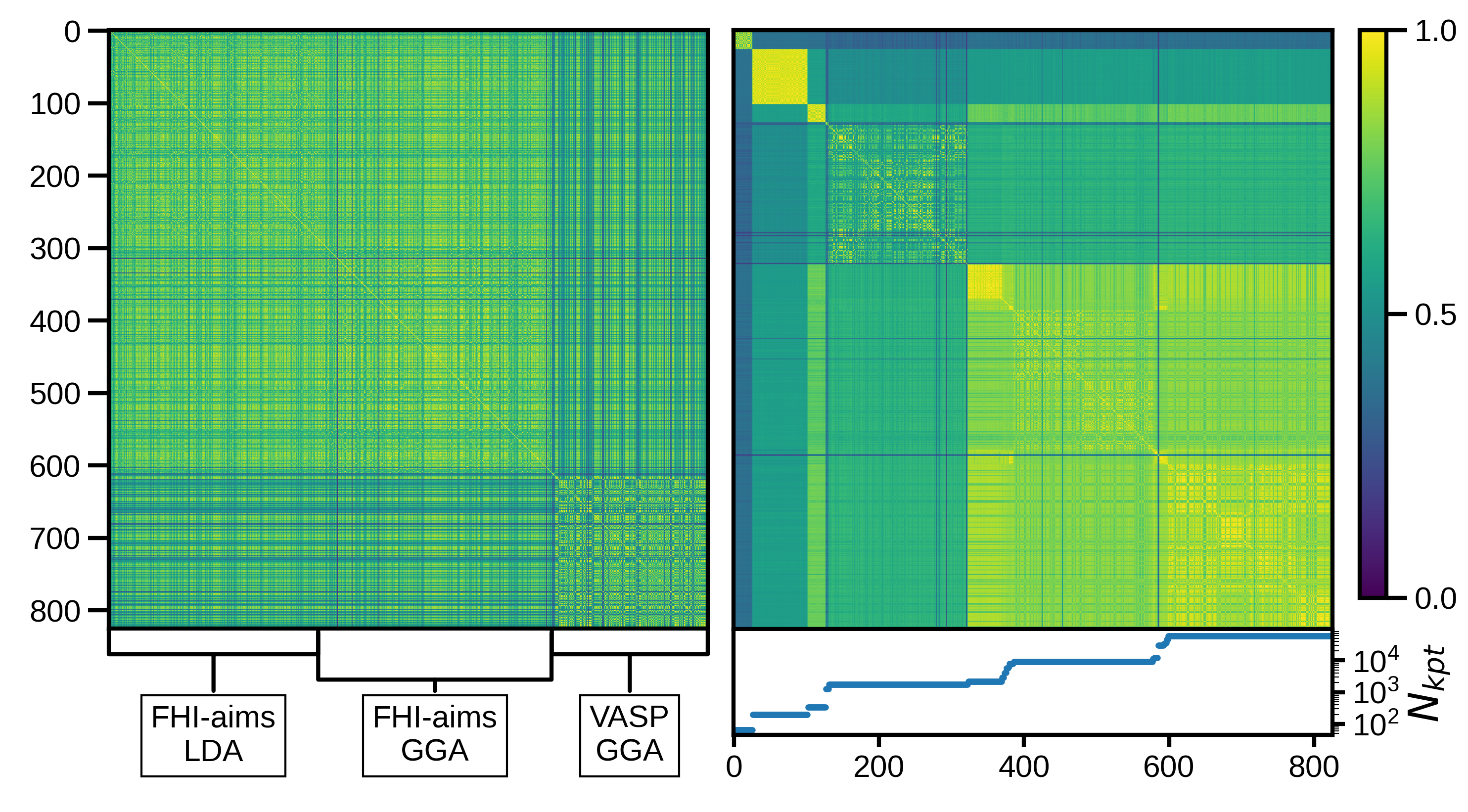}
        \caption{Similarity matrix of more than 800 individual calculations of fcc Al from the NOMAD Repository. The similarity coefficient is color-coded. On the left, the calculations are sorted with respect to the combination of DFT code and xc functional; on the right with respect to the total number of k-points, $N_{kpt}$, as indicated in the bottom. }
    \label{fig:codes}
\end{figure}

To reveal effects arising from computational parameters and tools, we study fcc Al. For this material,
we find more than 800 calculations from the NOMAD Repository \cite{NOMAD2018,NOMAD2019}. They have been carried out by different people with different codes and computational settings, depending on the individual purposes. Making use of the similarity coefficient we can gather useful information about the impact of these parameters on the DOS.  Again, in the ideal world, all calculations based on the same geometry should be identical. Figure \ref{fig:codes} shows the corresponding similarity matrix that shows that this is obviously not the case. On the left, the matrix is sorted according to the combination of DFT code and xc functional. Two clusters become apparent, basically distinguishing between two different codes. The one stemming from FHI-aims \cite{Blum2009} calculations contain both LDA and GGA results. Their similarity suggests that for fcc Al, the choice of the semi-local functional has no notable effect on the DOS. In contrast, there is overall less similarity between the VASP \cite{Kresse1996} and FHI-aims blocks. This behavior is rather unexpected and could be assigned to differences in the calculation of the DOS itself rather than in the electronic structure. Note that the pattern inside each block reflects all other differences in the calculations, like volume, {\bf k}-grid, computational parameters, etc.

The right panel of the figure shows the same data, now sorted by the total number of sampled {\bf k}-points, $N_{kpt}$. We see that the first about 350 entries are characterized by low similarity coefficients with all other data, with the exception of two smaller blocks that are formed by pure VASP and FHI-aims calculations, respectively. Above this index, the DOS of all calculations are very similar to each other, forming a significant cluster. We conclude that at this threshold, convergence with respect to  $N_{kpt}$ is reached. Like in the other example, here the patterns inside the bigger blocks stem from differences other than the {\bf k}-grid. This explains the dark lines inside the block and the two features mentioned above. 

To conclude from this example, our framework quantifies and visualizes the effect of computational parameters on the DOS, which allows us to identify outliers and helps in choosing representative parameters. We note in passing that the data in the NOMAD Repository are normalized, \ie brought to the same file formats and units, and having the same energy zero ($\mathrm{E_F}$ = 0\, eV). Nevertheless, they still reflect individual implementations. Thus, \eg applying additional smearing to the here presented data, enhances their similarity.

\subsection{Finding patterns in data}
\label{sec:patterns}

\begin{figure}[hbt]
    \centering
    \includegraphics[width = 0.65\textwidth]{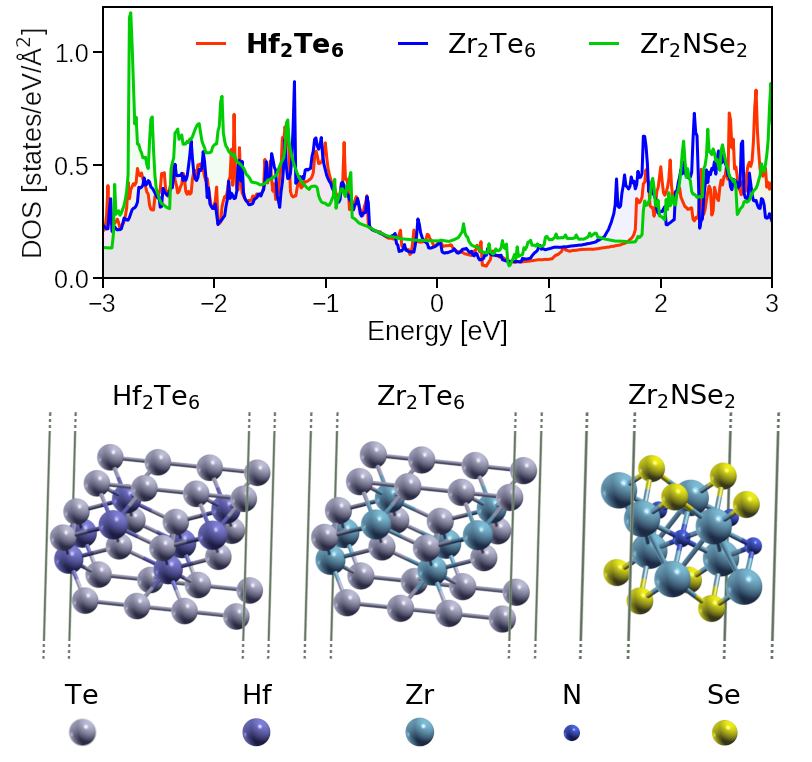}
        \caption{Cluster of materials from the from the C2DB \cite{Gjerding_2021} that exhibit a similar DOS, with Hf\fsub{2}Te\fsub{6} being the cluster center. Their crystal structures are displayed below. To enhance visibility, the unit cells are repeated in both in-plane directions.}
    \label{fig:clustering}
\end{figure}

Applied to well-curated datasets, the spectral fingerprint can be used to explore materials spaces in an unbiased manner when being combined with unsupervised learning. Adopting a clustering approach for the learning task, this has been recently demonstrated \cite{kuban2022} using data from the C2DB database \cite{Gjerding_2021}. Here, we provide another example of this kind to round off the potential of our descriptor and demonstrate how compact sets of materials can be found that are more similar to each other than a defined similarity threshold. The selected cluster was identified from a dataset consisting of 3491 materials. It comprises three materials, Hf\fsub{2}Te\fsub{6}, Zr\fsub{2}Te\fsub{6}, and Zr\fsub{2}NSe\fsub{2}. Their DOS and unit cells are depicted in Fig. \ref{fig:clustering}. Inspecting their crystal structures and electronic configurations, we see that this cluster exhibits expected as well as unexpected trends. The cluster center, \ie the material most similar to both others, is Hf\fsub{2}Te\fsub{6}. On the one hand, the latter and  Zr\fsub{2}Te\fsub{6} share the crystal symmetry (space group 59); also, Hf and Zr are isoelectronic. Thus their high similarity of $\mathrm{Tc} = 0.85$ in terms of the electronic structure could be anticipated. On the other hand, Zr\fsub{2}NSe\fsub{2} neither shares one nor the other characteristics with the other cluster members, nevertheless exhibits high similarity values of $\mathrm{Tc} = 0.79$ and $\mathrm{Tc} = 0.78$ with Hf\fsub{2}Te\fsub{6} and Zr\fsub{2}Te\fsub{6}, respectively. Thus, it can be considered as an {\it outlier}. Outliers are particularly interesting in view of finding candidate materials with specific features in materials spaces where one would not expect them. In fact, one of the great hopes is that data-driven analysis can guide researchers towards possibly interesting materials classes.

\section{Conclusions and outlook}
We have shown how a {\it spectral fingerprint} can be used in various conditions to measure the similarity of materials in terms of their electronic and spectroscopic properties. Our examples comprise a broad range of applications. We have, for instance, identified materials that share the electronic properties of their valence electrons or the DOS of one spin-component. Moreover, we could demonstrate that the same approach allows for assessing data quality. More specific, with the examples of SiC and PbI$_2$, it was illustrated how different treatment of exchange-correlation effects affect the electronic properties of semiconducting materials. More than 800 calculations of fcc Al were used to demonstrate the impact of computational parameters on the numerical precision of computed properties. Likewise, we probed the effect of structural features on the excitonic spectra in terms of the layer stacking in h-BN. All these examples not only allow us to determine the degree of interoperability of materials data but also help us to quantify more generally, how approximations, implementation details, computational parameters, or structural features affect calculations that are routinely carried out in numerous research groups all over the world. Automatizing such assessments would greatly enhance the understanding of computed results as they help analyzing many more materials calculations than what can be done in case-by-case investigations and single publications. The resulting findings will pave the way to incorporate physical reasoning into materials science data and annotate the data accordingly.

With the example of the noble metal silver, we also confronted various measured optical spectra with each other. Finally, we have shown how the spectral fingerprint can be combined with unsupervised learning to explore large data spaces and identify clusters of similar materials. 

What are the next steps for using and/or enhancing our tools? There are several open issues, and we address a few of them: 
\begin{itemize}
    \item Regarding the search for similar materials, one question concerns the choice of the reference calculation. What is a {\it representative} calculation? Ideally, one would choose (the) one carried out with the highest-level methodology, ensuring good convergence with respect to the computational parameters. Obviously, then also the calculations to compare with should fulfill the same criteria which in most cases would restrict the data space significantly. Thus, a milder criterion would be to use lower-level calculations as long as all calculations are from the same level of sophistication. To judge the impact of the method and parameters, in turn, the tools presented in Sections \ref{sec:methodology}, \ref{sec:parameters}, or \ref{sec:excitons} can guide the choice.  
    \item Obviously (see also Section \ref{sec:similar}), the similarity coefficient depends on where we put the focus on. Thus, having a certain function in mind, one could focus on certain energy regions or features. They might be as different as the band edges in semiconductors, the absorption onset in optical spectra, or the DOS at the Fermi level in metals or superconductors. The construction of the fingerprint is very flexible and prepared for all this.
    \item Methodology-wise we plan to consider additional similarity metrics. Let us exemplify the need with the example of excitons. Weakly-bound excitons are characterized by a redistribution of oscillator strength when compared to the independent-particle spectrum. This effect is well captured by the metric presented here. On the other hand, the rigid red shift of a stronly bound exciton is better captured by the Earth Mover's Distance \cite{Rubner1998}, also known as Wasserstein Distance. Therefore, a more complex metric is required that allows one to capture both types of excitonic effects or tell them apart.
\end{itemize}

To conclude, the here presented examples demonstrate that the spectral fingerprint can serve as a descriptor for a variety of scenarios. Besides the above described enhancements, for the future, we propose and plan to define similar criteria for other physical properties towards data analysis and quality assurance of computed and measured results. They could, for instance, bring to light differences in sample preparation, composition, dimensionality or other structural features, measurement conditions, and more. On the other hand, exploring interoperable data by unsupervised learning, will help finding either anticipated or unexpected trends in the data and may enable discoveries.

\section{Methods} \label{sec:methods}
In order to identify materials that share favorable properties in a large number of materials, a numeric representation, \ie a descriptor, of the investigated property is required. In this work, we focus on energy-dependent spectra that are here represented by a {\it spectral fingerprint}, analogous to the DOS fingerprint of Ref. \cite{kuban2022}. It consists of a two-dimensional raster image obtained through a special discretization of a spectrum, $s(E)$: First, the spectrum is converted to a histogram, $s_i$, with energy bins of variable width
\begin{eqnarray}
\Delta \varepsilon_i = n(\varepsilon_i,W,N)\, \Delta \varepsilon_{min},\label{eq:eps_i}
\end{eqnarray}
with $s_i$ being the integrated spectrum in the interval $\left[E_i,E_{i + 1} \right)$, with $E_i= E_{\mathrm{ref} }+ \varepsilon_i$, $E_{i + 1} = E_i+\Delta\varepsilon_i$, and $\varepsilon_0=0$. The integer-valued function $n(\varepsilon_i,W,N)$ is equal to $1$ for $\varepsilon_i=0$ and approaches $N$ monotonously for $|\varepsilon_i|>W$. Defined in this way, the integer $N\ge1$ and the real number $W > 0$ allow for increased resolution of the fingerprint in a region of width $W$ around $E_{\mathrm{ref}}$. If $N=1$, a uniform resolution is achieved. Finally, a grid of pixels is obtained by discretizing every column of the histogram in a grid of $N_{\rho}$ intervals of height $n(\varepsilon_i,W_H, N_H)\,\Delta \rho_{min}$. For the definition of the function $n$ and additional details on the fingerprint, see Ref. \cite{kuban2022}, where the spectral fingerprint is applied to the case of the electronic DOS.

To quantify the similarity between two spectra $A$ and $B$, we make use of the Tanimoto coefficient \cite{Willett1998}:
\begin{eqnarray}
\mathrm{Tc}(A,B) = \frac{A \cdot B}{A^2 + B^2 - A \cdot B},
\end{eqnarray}
which, for dichotomous descriptors like the spectral fingerprint, is a metric that ranges from 0 (not similar) to 1 (identical). For an overview of other similarity measures, \eg the Dice and cosine similarity, which are commonly applied in chemistry research, we refer to Ref. \cite{Willett1998}. For our specific application, we note that the Tanimoto coefficient has several advantages. First, the similarity is highly interpretable, as it can be understood as the intersection of the two DOS, divided by their union. Additionally, in contrast to, \eg the Dice coefficient, the Tanimoto coefficient obeys the triangle inequality, which is a favorable property for applications such as clustering. Furthermore, it is computationally cheap, as it can be calculated using only binary operations and bit counts, in contrast to, \eg the cosine similarity , which requires the calculation of a square root.

For the spectra shown in Fig. \ref{fig:similar_materials}, we use a non-uniform grid with $\Delta\varepsilon_\mathrm{min} = 0.05$, $\Delta\rho_\mathrm{min} \sim$0.001, $N = 21$, $N_H = 11$, $E_\mathrm{ref} = -2\;\mathrm{eV}$, $W = W_H = 7\;\mathrm{eV}$, $N_\rho = 256$, and a cutoff of $-10$ to $5\;\mathrm{eV}$. For the spectra in Fig. \ref{fig:LDA_GW}, we define the feature region by setting $E_\mathrm{ref} = 0\;\mathrm{eV}$,  $W = W_H = 10\;\mathrm{eV}$, $N_\rho = 512$, and the cutoff to $-10$ to $10\;\mathrm{eV}$. The fingerprints used to compare the optical spectra in Fig. \ref{fig:h-BN_spectra} employ uniform grids, \textit{i.e.} $N=1$, with $\Delta\varepsilon_\mathrm{min} = 0.02\;\mathrm{eV}$, $\Delta\rho_\mathrm{min} = 3 \times 10^{-3}$, $N_\rho = 256$, and a cutoff of $4$ to $8\;\mathrm{eV}$. The fingerprints used for Fig. \ref{fig:Al_experiment} use uniform grids with $N=1$, $\Delta\varepsilon_\mathrm{min} = 0.02\;\mathrm{eV}$, $\Delta\rho_\mathrm{min} = 1 \times 10^{-3}$, $N_\rho = 256$, and a cutoff of $2.2$ to $25\;\mathrm{eV}$.

\section{Data availability} \label{sec:data}
The data used in Section \ref{sec:similar} can be downloaded from the NOMAD Repository \cite{NOMAD2018,NOMAD2019} through the following links: GaAs \cite{NOMADGaAs}, GaP \cite{NOMADGaP}, NaBSi \cite{NOMADNaBSi}, Au \cite{NOMADAu}, CoFe \cite{NOMADCoFe}. The data used in Section \ref{sec:methodology} are calculated using the \exciting\ package \cite{Gulans2014} and are available in the NOMAD Repository \cite{NOMAD2018,NOMAD2019}. Specifically, the calculation for SiC can be accessed via \cite{NOMADSiCGW} and the dataset for PbI\fsub{2} via \cite{NOMADPbI2}. The data used in Section \ref{sec:excitons} can be accessed in the NOMAD Repository \cite{NOMAD2018,NOMAD2019} through \cite{Aggoune2018Data,Aggoune2022Data}. The calculations have been carried out with the \exciting\ package \cite{Gulans2014} by solving the Bethe-Salpeter equation of many-body perturbation theory (MBPT) in the linearized augmented planewave + local-orbital (LAPW+lo) basis \cite{Sagmeister2009,Vorwerk2019}. For details of the calculations, we refer to Ref. \cite{Aggoune2018}. For the experimental spectra analyzed in Section \ref{sec:experiment}, we refer to Refs. \cite{Ehrenreich1962,Robin1966,Hagemann1975,Leveque1983,Werner2009}. 
The data used in Section \ref{sec:parameters} are obtained from the NOMAD Repository \cite{NOMAD2018, NOMAD2019}. To access them, one has to query for calculations of materials that have been uploaded before Jan. 1, 2021, containing exclusively Al with space group number $225$, and having a DOS. An example of a search query returning this data can be found in Ref. \cite{NOMADAl}. 
The data stem from S. Curtarolo \textit{et al.} \cite{AFLOW, AFLOWAPI, AFLOWLIB.ORG}, C. Wolverton \textit{et al.} \cite{OQMD, Saal2013}, B. Bieniek \cite{NOMADBieniek}, C. Carbogno and B. Bieniek \cite{NOMADCarbogno}, O. Hofmann \cite{NOMADHofmann}, and K. Lejaeghere \cite{NOMADLejaeghere}. Additional calculations from B. Bieniek are not assigned a DOI. 
Our data analysis of Section \ref{sec:patterns} is based on materials from the high-throughput collection C2DB (Computational 2D Materials Database) \cite{Haastrup_2018, Gjerding_2021}. The specific data shown here can be accessed via the links provided in Refs. \cite{C2DBHf2Te6, C2DBZr2Te6, C2DBNSe2Zr2}.

\section{Acknowledgment}
This work received funding by the German Research Foundation (DFG) through the CRC 1404 (FONDA), project 414984028. Partial support from the NFDI consortium FAIRmat, DFG project 460197019, and the European Union’s Horizon 2020 research and innovation program under the grant agreement Nº 951786 (NOMAD CoE) is appreciated. The PbI$_2$ results have been carried out within the DFG priority program SPP2196 “Perovskite Semiconductors,” project Nr. 424709454. We thank Kathrin Glantschnig for providing the experimental spectra of Al, extracted from Refs. \cite{Ehrenreich1962,Robin1966,Hagemann1975,Leveque1983,Werner2009}.

\section{Conflicts of interest}
On behalf of all authors, the corresponding author states that there is no conflict of interest. 

\bibliographystyle{unsrt}
\bibliography{main}

\begin{thebibliography}{10}

\bibitem{MGI}
\url{https://mgi.gov/}.

\bibitem{MRS}
Isao Tanaka, Krishna Rajan, and Christopher Wolverton.
\newblock {Data-centric science for materials innovation}.
\newblock {\em MRS Bulletin}, 43(9):659–663, 2018.

\bibitem{Wilkinson2016}
M.~D.~Wilkinson et~al.
\newblock {The FAIR Guiding Principles for scientific data management and
  stewardship}.
\newblock {\em Scientific Data}, 3:160018, 2016.

\bibitem{Scheffler2022}
M.~Scheffler, M.~Aeschlimann, M.~Albrecht, T.~Bereau, H.-J. Bungartz,
  C.~Felser, M.~Greiner, A.~Gro\ss, C.~Koch, K.~Kremer, W.~E. Nagel,
  M.~Scheidgen, C.~Wöll, and Draxl C.
\newblock {FAIR data – new horizons for materials research}.
\newblock {\em Nature, in print}, 2022.

\bibitem{Lejaeghere2016}
Kurt Lejaeghere, Gustav Bihlmayer, Torbj{\"o}rn Bj{\"o}rkman, Peter Blaha,
  Stefan Bl{\"u}gel, Volker Blum, Damien Caliste, Ivano~E. Castelli, Stewart~J.
  Clark, Andrea Dal~Corso, Stefano de~Gironcoli, Thierry Deutsch, John~Kay
  Dewhurst, Igor Di~Marco, Claudia Draxl, Marcin Du{\l}ak, Olle Eriksson,
  Jos{\'e}~A. Flores-Livas, Kevin~F. Garrity, Luigi Genovese, Paolo Giannozzi,
  Matteo Giantomassi, Stefan Goedecker, Xavier Gonze, Oscar Gr{\r a}n{\"a}s,
  E.~K.~U. Gross, Andris Gulans, Fran{\c c}ois Gygi, D.~R. Hamann, Phil~J.
  Hasnip, N.~A.~W. Holzwarth, Diana Iu{\c s}an, Dominik~B. Jochym, Fran{\c
  c}ois Jollet, Daniel Jones, Georg Kresse, Klaus Koepernik, Emine K{\"u}{\c
  c}{\"u}kbenli, Yaroslav~O. Kvashnin, Inka L.~M. Locht, Sven Lubeck, Martijn
  Marsman, Nicola Marzari, Ulrike Nitzsche, Lars Nordstr{\"o}m, Taisuke Ozaki,
  Lorenzo Paulatto, Chris~J. Pickard, Ward Poelmans, Matt I.~J. Probert, Keith
  Refson, Manuel Richter, Gian-Marco Rignanese, Santanu Saha, Matthias
  Scheffler, Martin Schlipf, Karlheinz Schwarz, Sangeeta Sharma, Francesca
  Tavazza, Patrik Thunstr{\"o}m, Alexandre Tkatchenko, Marc Torrent, David
  Vanderbilt, Michiel~J. van Setten, Veronique Van~Speybroeck, John~M. Wills,
  Jonathan~R. Yates, Guo-Xu Zhang, and Stefaan Cottenier.
\newblock {Reproducibility in density functional theory calculations of
  solids}.
\newblock {\em Science}, 351(6280):aad3000, 2016.

\bibitem{Gulans2018}
Andris Gulans, Anton Kozhevnikov, and Claudia Draxl.
\newblock {Microhartree precision in density functional theory calculations}.
\newblock {\em Phys. Rev. B}, 97:161105, Apr 2018.

\bibitem{Jensen2017}
Stig~Rune Jensen, Santanu Saha, José~A. Flores-Livas, William Huhn, Volker
  Blum, Stefan Goedecker, and Luca Frediani.
\newblock {The Elephant in the Room of Density Functional Theory Calculations}.
\newblock {\em The Journal of Physical Chemistry Letters}, 8(7):1449--1457,
  2017.
\newblock PMID: 28291362.

\bibitem{Nabok2016}
Dmitrii Nabok, Andris Gulans, and Claudia Draxl.
\newblock {Accurate all-electron ${G}_{0}{W}_{0}$ quasiparticle energies
  employing the full-potential augmented plane-wave method}.
\newblock {\em Phys. Rev. B}, 94:035118, Jul 2016.

\bibitem{Rangel2020}
Tonatiuh Rangel, Mauro {Del Ben}, Daniele Varsano, Gabriel Antonius, Fabien
  Bruneval, Felipe~H. {da Jornada}, Michiel~J. {van Setten}, Okan~K. Orhan,
  David~D. O’Regan, Andrew Canning, Andrea Ferretti, Andrea Marini,
  Gian-Marco Rignanese, Jack Deslippe, Steven~G. Louie, and Jeffrey~B. Neaton.
\newblock {Reproducibility in G0W0 calculations for solids}.
\newblock {\em Computer Physics Communications}, 255:107242, 2020.

\bibitem{Ramprasad2017}
Rampi Ramprasad, Rohit Batra, Ghanshyam Pilania, Arun Mannodi-Kanakkithodi, and
  Chiho Kim.
\newblock {Machine learning in materials informatics: recent applications and
  prospects}.
\newblock {\em npj Computational Materials}, 3(1):54, 2017.

\bibitem{Isayev2015}
Olexandr Isayev, Denis Fourches, Eugene~N. Muratov, Corey Oses, Kevin Rasch,
  Alexander Tropsha, and Stefano Curtarolo.
\newblock {Materials Cartography: Representing and Mining Materials Space Using
  Structural and Electronic Fingerprints}.
\newblock {\em Chemistry of Materials}, 27(3):735--743, 2015.

\bibitem{Mahmoud2020}
Chiheb Ben~Mahmoud, Andrea Anelli, G\'abor Cs\'anyi, and Michele Ceriotti.
\newblock {Learning the electronic density of states in condensed matter}.
\newblock {\em Phys. Rev. B}, 102:235130, Dec 2020.

\bibitem{Gjerding_2021}
Morten~Niklas Gjerding, Alireza Taghizadeh, Asbj{\o}rn Rasmussen, Sajid Ali,
  Fabian Bertoldo, Thorsten Deilmann, Nikolaj~R{\o}rb{\ae}k Kn{\o}sgaard, Mads
  Kruse, Ask~Hjorth Larsen, Simone Manti, Thomas~Garm Pedersen, Urko
  Petralanda, Thorbj{\o}rn Skovhus, Mark~Kamper Svendsen, Jens~J{\o}rgen
  Mortensen, Thomas Olsen, and Kristian~Sommer Thygesen.
\newblock {Recent progress of the Computational 2D Materials Database (C2DB)}.
\newblock {\em 2D Materials}, 8(4):044002, jul 2021.

\bibitem{Knosgaard2022}
Nikolaj~R{\o}rb{\ae}k Kn{\o}sgaard and Kristian~Sommer Thygesen.
\newblock {Representing individual electronic states for machine learning GW
  band structures of 2D materials}.
\newblock {\em Nature Communications}, 13(1):468, 2022.

\bibitem{kuban2022}
Martin Kuban, Santiago Rigamonti, Markus Scheidgen, and Claudia Draxl.
\newblock {Density-of-states similarity descriptor for unsupervised learning
  from materials data}, preprint: \url{https://arxiv.org/abs/2201.02187}, 2022.

\bibitem{De2016}
Sandip De, Albert~P. Bartók, Gábor Csányi, and Michele Ceriotti.
\newblock {Comparing molecules and solids across structural and alchemical
  space}.
\newblock {\em Phys. Chem. Chem. Phys.}, 18:13754--13769, 2016.

\bibitem{Willett1998}
Peter Willett, John~M. Barnard, and Geoffrey~M. Downs.
\newblock {Chemical Similarity Searching}.
\newblock {\em Journal of Chemical Information and Computer Sciences},
  38(6):983--996, 1998.

\bibitem{NOMAD2018}
Claudia Draxl and Matthias Scheffler.
\newblock {NOMAD: The FAIR Concept for Big-Data-Driven Materials Science}.
\newblock {\em MRS Bulletin}, 43:676, 2018.

\bibitem{NOMAD2019}
Claudia Draxl and Matthias Scheffler.
\newblock The {NOMAD} laboratory: from data sharing to artificial intelligence.
\newblock {\em JPM}, 2(3):036001, 2019.

\bibitem{NOMADSiCGW}
\url{https://nomad-lab.eu/entry/id/kUqd_BDKGmTWhJcbvy_puqwA9vbi}.

\bibitem{Vona2021}
Cecilia Vona, Dmitrii Nabok, and Claudia Draxl.
\newblock {Electronic Structure of (Organic-)Inorganic Metal Halide
  Perovskites: The Dilemma of Choosing the Right Functional}.
\newblock {\em Advanced Theory and Simulations}, 5(1):2100496, 2022.

\bibitem{Gulans2014}
Andris Gulans, Stefan Kontur, Christian Meisenbichler, Dimitrii Nabok, Pasquale
  Pavone, Santiago Rigamonti, Stephan Sagmeister, Ute Werner, and Claudia
  Draxl.
\newblock {{\usefont{T1}{lmtt}{b}{n}exciting}: a full-potential all-electron
  package implementing density-functional theory and many-boyd perturbation
  theory}.
\newblock {\em J. Phys.: Condens. Matter}, 26(36):363202, 2014.

\bibitem{Gahailler1969}
Ch. G\"ahwiller and G.~Harbeke.
\newblock {Excitonic Effects in the Electroreflectance of Lead Iodide}.
\newblock {\em Phys. Rev.}, 185(1141), Sep 1969.

\bibitem{Ahuja2002}
R.~Ahuja, H.~Arwin, A.~Ferreira Da~Silva, C.~Persson, J.~M.
  Osorio-Guill{\'{e}}n, J.~Souza De~Almeida, C.~Moyses Araujo, E.~Veje,
  N.~Veissid, C.~Y. An, I.~Pepe, and B.~Johansson.
\newblock {Electronic and optical properties of lead iodide}.
\newblock {\em J. Appl. Phys.}, 92(7219), 12 2002.

\bibitem{Shen2018}
Chenhai Shen and Guangtao Wang.
\newblock {Electronic and optical properties of bilayer PbI2: A
  first-principles study}.
\newblock {\em J. Phys. D: Appl. Phys.}, 51(035301), 1 2018.

\bibitem{Aggoune2018}
Wahib Aggoune, Caterina Cocchi, Dmitrii Nabok, Karim Rezouali, Mohamed~Akli
  Belkhir, and Claudia Draxl.
\newblock {Dimensionality of excitons in stacked van der Waals materials: The
  example of hexagonal boron nitride}.
\newblock {\em Phys. Rev. B}, 97:241114(R), 2018.

\bibitem{Ehrenreich1962}
H.~Ehrenreich and H.~R. Philipp.
\newblock {\em Phys. Rev.}, 128:1622, 1962.

\bibitem{Robin1966}
S.~Robin.
\newblock {Propri{\'e}t{\'e}s optiques de l’argent et du palladium dans
  l’ultraviolet lointain}, 1966.

\bibitem{Hagemann1975}
H.~J. Hagemann, W.~Gudat, and C.~Kunz.
\newblock {\em Phys. Rev. B}, 65:742, 1975.

\bibitem{Leveque1983}
G.~Leveque, C.~G. Olson, and D.~W. Lynch.
\newblock {\em Phys. Rev. B}, 27:4654, 1983.

\bibitem{Werner2009}
W.~S.~M. Werner, K.~Glantschnig, and C.~Ambrosch-Draxl.
\newblock {\em J. Phys. Chem. Ref. Data}, 38:1013, 2009.

\bibitem{Blum2009}
Volker Blum, Ralf Gehrke, Felix Hanke, Paula Havu, Ville Havu, Xinguo Ren,
  Karsten Reuter, and Matthias Scheffler.
\newblock {Ab initio molecular simulations with numeric atom-centered
  orbitals}.
\newblock {\em Computer Physics Communications}, 180(11):2175--2196, 2009.

\bibitem{Kresse1996}
G.~Kresse and J.~Furthm\"uller.
\newblock {Efficient iterative schemes for ab initio total-energy calculations
  using a plane-wave basis set}.
\newblock {\em Phys. Rev. B}, 54:11169--11186, Oct 1996.

\bibitem{Rubner1998}
Y.~Rubner, C.~Tomasi, and L.J. Guibas.
\newblock A metric for distributions with applications to image databases.
\newblock In {\em Sixth International Conference on Computer Vision (IEEE Cat.
  No.98CH36271)}, pages 59--66, 1998.

\bibitem{NOMADGaAs}
\url{https://nomad-lab.eu/entry/id/zkkMIAPyn4OCbdEdW21DZTeretQ3}.

\bibitem{NOMADGaP}
\url{https://nomad-lab.eu/entry/id/yMEbPhw-ttwsWZEsEVycez647OKH}.

\bibitem{NOMADNaBSi}
\url{https://nomad-lab.eu/entry/id/mFXpw4Tn8E1iJBs1-nkeWqcBR9sf}.

\bibitem{NOMADAu}
\url{https://nomad-lab.eu/entry/id/sYp4jkDZtVwzFyc7WN0cL5RCFq1s}.

\bibitem{NOMADCoFe}
\url{https://nomad-lab.eu/entry/id/m2EbbBqduN-MIEh7U3kaA00_62ic}.

\bibitem{NOMADPbI2}
\url{https://dx.doi.org/10.17172/NOMAD/2021.10.26-1}.

\bibitem{Aggoune2018Data}
\url{http://dx.doi.org/10.17172/NOMAD/2018.06.05-1}.

\bibitem{Aggoune2022Data}
\url{https://dx.doi.org/10.17172/NOMAD/2022.01.23-1}.

\bibitem{Sagmeister2009}
Stephan Sagmeister and Claudia Ambrosch-Draxl.
\newblock {Time-Dependent Density Functional Theory versus Bethe--Salpeter
  equation: An All-Electron Study}.
\newblock {\em PCCP}, 11(22):4451--4457, 2009.

\bibitem{Vorwerk2019}
Christian Vorwerk, Benjamin Aurich, Caterina Cocchi, and Claudia Draxl.
\newblock {Bethe{\textendash}Salpeter equation for absorption and scattering
  spectroscopy: implementation in the exciting code}.
\newblock {\em Electronic Structure}, 1(3):037001, aug 2019.

\bibitem{NOMADAl}
\url{https://nomad-lab.eu/prod/rae/gui/search?visualization=elements&dft.searchable_quantities=electronic_dos&dft.spacegroup=225&until_time=2021-01-01T00%3A00%3A00.000Z&only_atoms=Al}.

\bibitem{AFLOW}
\url{http://aflowlib.org/}.

\bibitem{AFLOWAPI}
Richard~H. Taylor, Frisco Rose, Cormac Toher, Ohad Levy, Kesong Yang, Marco
  {Buongiorno Nardelli}, and Stefano Curtarolo.
\newblock {A RESTful API for exchanging materials data in the AFLOWLIB.org
  consortium}.
\newblock {\em Computational Materials Science}, 93:178--192, 2014.

\bibitem{AFLOWLIB.ORG}
Stefano Curtarolo, Wahyu Setyawan, Shidong Wang, Junkai Xue, Kesong Yang,
  Richard~H. Taylor, Lance~J. Nelson, Gus~L.W. Hart, Stefano Sanvito, Marco
  Buongiorno-Nardelli, Natalio Mingo, and Ohad Levy.
\newblock {AFLOWLIB.ORG: A distributed materials properties repository from
  high-throughput ab initio calculations}.
\newblock {\em Computational Materials Science}, 58:227--235, 2012.

\bibitem{OQMD}
\url{http://oqmd.org/}.

\bibitem{Saal2013}
James~E. Saal, Scott Kirklin, Muratahan Aykol, Bryce Meredig, and C.~Wolverton.
\newblock {Materials Design and Discovery with High-Throughput Density
  Functional Theory: The Open Quantum Materials Database (OQMD)}.
\newblock {\em JOM}, 65(11):1501--1509, Nov 2013.

\bibitem{NOMADBieniek}
\url{https://dx.doi.org/10.17172/NOMAD/2017.01.24-1}.

\bibitem{NOMADCarbogno}
\url{https://dx.doi.org/10.17172/NOMAD/2020.07.27-1}.

\bibitem{NOMADHofmann}
\url{https://dx.doi.org/10.17172/NOMAD/2020.07.29-1}.

\bibitem{NOMADLejaeghere}
\url{https://dx.doi.org/10.17172/NOMAD/2016.10.19-1}.

\bibitem{Haastrup_2018}
Sten Haastrup, Mikkel Strange, Mohnish Pandey, Thorsten Deilmann, Per~S
  Schmidt, Nicki~F Hinsche, Morten~N Gjerding, Daniele Torelli, Peter~M Larsen,
  Anders~C Riis-Jensen, Jakob Gath, Karsten~W Jacobsen, Jens~J{\o}rgen
  Mortensen, Thomas Olsen, and Kristian~S Thygesen.
\newblock {The Computational {2D} Materials Database: high-throughput modeling
  and discovery of atomically thin crystals}.
\newblock {\em 2D Materials}, 5(4):042002, sep 2018.

\bibitem{C2DBHf2Te6}
\url{https://cmrdb.fysik.dtu.dk/c2db/row/Hf2Te6-3f5c06f4bf18}.

\bibitem{C2DBZr2Te6}
\url{https://cmrdb.fysik.dtu.dk/c2db/row/Zr2Te6-8ef6448a7da4}.

\bibitem{C2DBNSe2Zr2}
\url{https://cmrdb.fysik.dtu.dk/c2db/row/NSe2Zr2-d567fba5f5ba}.

\end{thebibliography}

\end{document}